\documentclass[11pt]{article}
\usepackage[T1]{fontenc}
\usepackage[utf8]{inputenc}
\usepackage{newpxtext}  
\usepackage{bm,amsmath,amssymb,mathtools,graphicx}
\usepackage{textcomp,gensymb,units,upquote}
\usepackage[a4paper]{geometry}
\usepackage{epsfig}
\usepackage[parfill]{parskip}
\usepackage[matrix,arrow,color]{xy}
\usepackage[usenames]{color}
\usepackage{mathrsfs}
\usepackage[font={small}]{caption}
\usepackage{subcaption}
\usepackage[export]{adjustbox}
\usepackage{float}
\usepackage{wrapfig}
\usepackage{microtype}
\usepackage{slashed}
\usepackage{braket}
\usepackage[pdftex]{hyperref}

\input{xy}
\xyoption{all}

\input{epsf}

\makeatletter

\usepackage{setspace}

\usepackage{lscape}

\catcode`@=11
\renewcommand{\section}
{\@startsection{section}{1}{0pt}{\medskipamount}{\medskipamount}{\large\bf}}
\makeatletter\renewcommand{\subsection}
{\@startsection{subsection}{2}{\z@}{-3.25ex plus -1ex minus -.2ex}
{1.5ex plus .2ex}{\it }}

\numberwithin{equation}{section}
\catcode`@=12

\newcommand{\ban}{\begin{eqnarray}}
\newcommand{\ean}{\end{eqnarray}}

\newcommand{\Tr}{{\rm Tr}}
\newcommand{\tr}{{\rm tr}}

\newcommand{\cS}{{\cal S}}
\newcommand{\cB}{{\cal B}}

\newcommand{\cH}{{\cal H}}
\newcommand{\cA}{{\cal A}}

\newcommand{\cO}{{\cal O}}

\newcommand{\cR}{{\cal R}}

\newcommand{\sfe}{{\mathsf{e}}}
\newcommand{\sfa}{{\mathsf{a}}}
\newcommand{\sfb}{{\mathsf{b}}}

\newcommand{\sfx}{{\mathsf{x}}}
\newcommand{\sfy}{{\mathsf{y}}}

\newcommand{\complex}{{\mathbb C}} 
\newcommand{\real}{{\mathbb R}} 

\def\e{{\,\rm e}\,}

\def\ii{{\,{\rm i}\,}}
\def\dd{{\rm d}}

\def\beq{\begin{equation}}
\def\bee{\begin{equation}}
\def\eeq{\end{equation}}
\def\bea{\begin{eqnarray}}
\def\eea{\end{eqnarray}}
\def\bd{\begin{displaymath}}
\def\ed{\end{displaymath}}

\newcommand{\Cint}{\int\kern-10.5pt-\kern7pt}

\newcommand{\be}{\begin{equation}}
\newcommand{\ee}{\end{equation}}
\newcommand{\bal}{\begin{align}}
\newcommand{\eal}{\end{align}}

\newcommand\fverbit{\egroup\item[\fbox{\unhbox\pippobox}]}
\newbox\pippobox

{

\def\be{\begin{equation}}
\def\ee{\end{equation}}
\def\bea{\begin{eqnarray}}
\def\eea{\end{eqnarray}}

\begin{document}

\begin{titlepage}
\setcounter{page}{1}

\vskip 5cm

\begin{center}

\vspace*{3cm}

{\Large FLUCTUATION THEOREMS, QUANTUM CHANNELS \\[8pt] AND \\[8pt] GRAVITATIONAL ALGEBRAS}

\vspace{15mm}

{\large Michele Cirafici}
\\[6mm]
\noindent{\em Dipartimento di Matematica, Informatica e Geoscienze, \\ Universit\`a di Trieste, Via A. Valerio 12/1, I-34127,
 \\ Institute for Geometry and Physics \& INFN, Sezione di Trieste,  Trieste, Italy 
}\\ Email: \ {\tt michelecirafici@gmail.com}

\vspace{15mm}

\begin{abstract}
\noindent In this note we study nonequilibrium fluctuations in gravitational algebras within de Sitter space. An essential aspect of this study is quantum measurement theory, which allows us to access the dynamical fluctuations of observables via a two-point measurement scheme. Using this formalism, we establish specific fluctuation theorems. Additionally, we demonstrate that quantum channels are represented by subfactors, using the relationship between measurement theory and quantum channels. We also comment on implementing a quantum channel using Jones' theory of subfactors.
\end{abstract}

\vspace{15mm}

\today

\end{center}
\end{titlepage}


\tableofcontents


\vspace{0.5cm}

\begin{flushright}
 \textit{Caminantes, no hay caminos, hay que caminar...} \\ Anonymous
\end{flushright}
\vspace{0.5cm}

\section{Introduction}

In ordinary quantum mechanics, a physical system and an observer are two separate entities. They interact when the observer performs a measurement, resulting in the system's state collapsing according to the Born rule. However, the situation changes in the presence of gravity. Since the observer also gravitates, they cannot be considered completely decoupled from the quantum system.

Recent research has shown that incorporating perturbative quantum gravity effects can significantly impact quantum systems \cite{Witten:2021unn}. The algebra of observables transitions from a type $\mathrm{III}$ von Neumann algebra to a type $\mathrm{II}$ algebra. This type of algebra supports density matrices and traces, and therefore one can define an entropy, but lacks irreducible representations or pure states. It is well-suited for understanding perturbative quantum gravity as a coarse-grained theory, akin to thermodynamics, where information theory quantifies our ignorance of microscopic details.

In spacetimes with an asymptotic boundary, an observer at infinity can gravitationally dress observables. However, this is not possible in a closed universe, such as de Sitter space, where spatial slices are closed manifolds. As emphasized in \cite{Chandrasekaran:2022cip,Witten:2023qsv}, an observer is required to impose gravitational constraints properly. The resulting algebra of observables is constructed from the type $\mathrm{III}$ algebra of quantum field observables on the static patch by introducing an observer and imposing gravitational constraints, resulting in a type $\mathrm{II}_1$ algebra. This perspective clarifies several results. For instance, empty de Sitter space corresponds to a state of maximal entropy, explaining the thermal nature of correlators. Thermal fluctuations can be interpreted as entropic fluctuations. Similarly, in a non-empty universe, the entropy of a semi-classical state is given by the generalized entropy. Additional discussions on the relationship between observers and gravitational algebras can be found in \cite{AliAhmad:2024wja,AliAhmad:2023etg,Chen:2024rpx,DeVuyst:2024pop,Faulkner:2024gst,fewster,Gomez:2023jbg,Gomez:2023upk,Hoehn:2023ehz,Jensen:2023yxy,Kudler-Flam:2024psh,Kudler-Flam:2023qfl,Witten:2023xze,Xu:2024hoc}. For applications to black holes and other setups, see \cite{AliAhmad:2024eun,Boruch:2024kvv,Cirafici:2024jdw,Chandrasekaran:2022eqq}.


Note that a more complete theory should derive, rather than assume, the presence of the observer. It is worth mentioning that there are attempts to obtain a non-trivial set of physical states without introducing an observer, but instead through a group-averaging procedure. Indeed as shown in \cite{Higuchi:1991tk,Higuchi:1991tm}, one can construct de Sitter-invariant states with infinite norm by smearing over the de Sitter group and then construct a finite inner product on these states by dividing by the volume of the group.

In line with the thermodynamic coarse-grained interpretation, this note addresses the dynamical fluctuations associated with certain states. Ordinary quantum mechanics provides a straightforward method for understanding static fluctuations of observables: the experimenter prepares several identical copies of the same system and performs projective measurements of an observable $X$, obtaining the probability distribution of the eigenvalues of $X$ in a particular state $\rho$. To study dynamical fluctuations, one must allow the system to evolve for some time after the initial measurement before performing a second measurement. This approach is known as the two-point or two-time measurement scheme.

In this note, we aim to explore thermodynamic fluctuations of observables in perturbative quantum gravity using a two-time measurement scheme. Our primary motivation is to understand certain nonequilibrium aspects of the dynamics, following our previous work \cite{Cirafici:2024jdw}. We will establish general fluctuation theorems for physical quantities, extending the results of Jarzynski \cite{jarzynski} and England \cite{england} to perturbative quantum gravity in de Sitter space.

Considering measurements in de Sitter space naturally leads to the study of quantum channels, which can be seen as generalized measurements without recording the outcome \cite{preskill}. Mathematically, a quantum channel is a trace-preserving, completely positive map. The theory of type $\mathrm{II}_1$ factors includes natural trace-preserving completely positive maps, such as the conditional expectation map, which identifies a subfactor. We focus on finite index subfactors, where we provide a physical interpretation of the structures relevant to their classification. Notably, for infinite-dimensional algebras, an observer may require a hierarchy of auxiliary vector spaces to set up quantum channels properly, unlike generalized quantum measurements, where a single "ancilla" space suffices.

While finalizing this submission, we received \cite{vanderHeijden:2024tdk}, where Jones' basic construction for a type $\mathrm{II}_1$ factor is also discussed. In \cite{vanderHeijden:2024tdk}, this construction is applied to a model of evaporating black holes to address the black hole information problem. In this note, however, we focus on the hyperfinite type $\mathrm{II}_1$ factor, which describes physics in the static patch of de Sitter space, and relate the associated Jones construction to quantum channels. The two perspectives seem complementary and compatible; however it would be interesting to further investigate their relationship.

\section{Quantum Measurement Theory} \label{QMT}

In this Section we briefly review certain aspects of quantum measurement theory in finite dimensions, including two-times measurements and their relation to nonequilibrium physics, and quantum channels. We will follow the texts \cite{preskill,strasberg}.

\paragraph{Projective measurements.} Assume we have a finite dimensional Hilbert space $\cH$ and consider the algebra of bounded operators $\cB (\cH)$. Consider a set of orthogonal projection operators $\{ \pi_s \}$, $\pi_s \pi_r = \delta_{sr} \pi_s$ and $\sum_{s=1}^n \pi_s = \mathbf{1}$. We write the spectral decomposition of an observable $O$ as $O = \sum_s \lambda_s \pi_s$.

Assume that the system is initially in a state described by the density matrix $\rho$. The probability of the $s^{\mathrm{th}}$ outcome is then
$
p_s = \tr \, \rho \, \pi_s
$.
After having measured the outcome $s$ the system is in the state
\be
\rho'_s = \frac{\pi_s \, \rho \, \pi_s}{p_s} \, .
\ee
If we don't record the measurement outcome, we have to average over all possible post-measurement states weighted by their probability. The average post-measurement state (that is unconditional to the measure of the outcome $s$)
\be \label{no-out}
\rho' = \sum_s p_s \rho_s' = \sum_s \pi_s \, \rho \, \pi_s
\ee
describes the averaged effect of a quantum measurement. Note that since we haven't recorded the outcome the probabilities $p_s$ have canceled. 

\paragraph{Ancillas.} These expressions can also be derived from a unitary dynamics if we introduce an auxiliary $n$ dimensional quantum system, the ancilla $A$. One assumes that originally the ancilla is in a pure state $\rho_A = \ket{1}\bra{1}_A$. 
One postulates that the interaction between the ancilla and the original system is given by the operator
\be
V = \sum_{s=1}^n \pi_s \otimes \sum_{r=1}^n \ket{r+s-1}\bra{r} \, .
\ee
Note that we are assuming the ancilla has a basis of states of the same cardinality as the projectors of the original system. Ordering of labels is cyclic. It is easy to see that $V$ is a unitary operator. The effect of the interaction between the system and the ancilla is that the combined system is described by
$
\rho_{AS}' = V \rho \otimes \rho_A V^\dagger
$,
so that tracing over the ancilla degrees of freedom
\be
\rho' = \tr_A V \rho \otimes \rho_A V^\dagger = \sum_{s=1}^n \pi_s \, \rho \, \pi_s
\ee
reproduces the effect of a projective measurement. Similar considerations hold for $\rho_s'$. Arguments along these lines can be used to describe also generalized measurements.

\paragraph{Quantum Channels.}

An experimenter can manipulate a quantum system in many ways, without necessarily measuring its state. The density matrix of a state can in general evolve according to
\be \label{qchannel}
\rho \longrightarrow \rho' = \sum_k E_k \, \rho \, E_k^\dagger \, ,  \qquad \sum_k E^\dagger_k \, E_k = 1 \, ,
\ee
where the operators $E_k$ are called Kraus operators. This operation is called a quantum channel. It generalizes ordinary unitary evolution, which is recovered when there is only one Kraus operator. It is the most general control operation that an experimental can perform on the system. Since the expression \eqref{qchannel} is formally the same as \eqref{no-out}, quantum channels can be seen as (generalized) measurements where the outcome is not recorded \cite{preskill}.

Consistency requires this operation to satisfy two conditions: it has to be trace-preserving (that is if $\Tr \rho = 1$ then $\Tr \rho' = 1$) and completely positive (a map $\mathcal{C}$ is positive is it sends positive elements to positive elements; 
it is completely positive if its tensor product with the identity map in $\cB (\complex^n)$ is positive for every $n$; physically the auxiliary Hilbert space ensures that quantum channels map states to states also when they are acting on a part of the system). 

Consider for example a system described by a state $\rho$. We act with the Kraus operator $E_k$ to obtain the state
\be
\frac{E_k \rho E_k^\dagger}{\Tr E_k \rho E_k^\dagger}
\ee
with probability $p_k = \Tr E_k \rho E_k^\dagger$. The expectation value of any observable $A$ conditioned on a generalized measurement described by the set of Kraus operators $\{ E_k \}$ is obtained by summing over all possible outcomes, weighted by their probability $p_k$; that is
\be \label{CEandK}
\mathbb{E} \left( A \vert \rho , \{ E_k \} \right) = \Tr \left( A \sum_k E_k \rho E_k^\dagger \right) \, .
\ee
In other words we can equivalently describe a quantum channel as a conditional expectation. 

In the following we will find that there are natural trace-preserving completely positive maps in the context of type $\rm{II}_1$ which also can be described as conditional expectations.  

\paragraph{Two-time measurements and nonequilibrium dynamics.}

A two-time measurement is characterized by the fact that the quantum system is let free to evolve in time between the two measurements. Assume that our system is initially at time $t=0$ in a state described by the density matrix $\rho$. Consider an observable $X$ with spectral decomposition $X = \sum_x x \, \Pi_x$. Assume that an observer performs a projective measurement of $X$ and finds the value $x$. As explained above now the system is in the state $\Pi_x \rho \Pi_x / p(x)$. We now let the system evolve in time with an Hamiltonian $H$ and perform a second measurement at the time $t>0$. The probability of obtaining $x'$ having already measured $x$ is
\be
p (x' \vert x) = \frac{1}{p(s)} \tr \left( \Pi_{x'} \e^{- \ii t H} \Pi_x \rho \Pi_x \e^{\ii t H} \right)
\ee
so that the probability of obtaining the two values $x$ and $x'$ is
\be
p (x' , x) =  \tr \left( \Pi_{x'} \e^{- \ii t H} \Pi_x \rho \Pi_x \e^{\ii t H} \right)
\ee
Note that if we assume that projections correspond to pure states $\Pi_x = \ket{x}\bra{x}$, then the above formulas read
\begin{align}
p (x' \vert x) &= \left\vert \braket{x' \vert U_{t,0} \vert x } \right\vert^2 \cr
p (x' , x) &= | \braket{x' \vert U_{t,0} \vert x } \vert^2  \braket{x \vert \rho \vert x}
\end{align}
In particular $p (x' \vert x) $ has also the interpretation of the transition probability between two states. We will also loosely refer to $p (x' , x)$ as a transition probability.

One can obtain information about the nonequilibrium dynamics comparing a process with its time-reverse process, with probability $p^{tr} (x,x')$. If one is interested in a certain function $f$ of the outcome and set $\Delta f = f (x') - f (x)$, then one can prove that \cite{strasberg}
\be
\frac{p (\Delta f)}{p^{tr} (-\Delta f)} = \e^{\Delta f}
\ee
One can chose $\Delta f = \beta (\Delta E - \Delta F)$ where $\Delta E$ is the change in energy of the system obtained from the two measurements, and $\Delta F$ is a constant associated with the system free energy (the logarithm of the normalization of the density matrix). If one interprets the change of energy as work $\Delta E  =  w$ one finds
\be
\frac{p (w)}{p^{tr} (-w)} = \e^{\beta (w - \Delta F)}
\ee
which averaged gives the quantum analog of the Jarzynski's equality \cite{jarzynski}
\be
\braket{\e^{- \beta w}} = \e^{- \beta \Delta F}
\ee
We refer the reader to \cite{strasberg} for a more in depth discussion and a more complete overview of the relevant literature.

\section{Observers and gravitational algebras}

We will now focus on aspects of gravitational algebras in de Sitter space. Specifically, as stressed in \cite{Chandrasekaran:2022cip}, the algebra of operators in de Sitter, being a closed universe, requires an observer to be operationally defined. If we consider the static patch accessible to the observer, then the algebra of observables $\cA_0$ can be defined as the algebra generated by quantum fields along the wordline of the observer. This algebra acts on a ``code'' subspace of the Hilbert space $\cH_0$. The physical algebra of observables in the static patch is then obtained adding to considering $\cA_0$ and adding the information about the observer.

If $H$ is the Hamiltonian which generates time translations on the static patch, adding the observer gives the new Hamiltonian
\be
\widehat{H} = H + H_{\mathrm{obs}} = H + q \, ,
\ee
where we have adapted the simplest model of an observer as in \cite{Chandrasekaran:2022cip}, a simple clock whose energy is bounded from below. The fact that there interactions between the observer and the quantum fields can be neglected is equivalent to the limit $G_N \longrightarrow 0$. The full Hilbert space is now
\be
\cH = \cH_0 \otimes L^2 (\real_+) \, .
\ee 
The physical algebra is obtained by imposing the Hamiltonian constraint
\be
\cA = \left( \cA_0 \otimes \cB (L^2 (\real_+)) \right)^{\widehat{H}} \, .
\ee
This algebra is obtained through a two-step procedure. First, one takes the crossed product of the algebra $\cA_0$ with the one-parameter group of automorphisms generated by $H$, resulting in a $\mathrm{II}_\infty$ algebra. Next, the observer’s energy is constrained to be bounded from below by applying the appropriate projection $\Theta$ (which is $1$ for $q \ge 0$ and zero otherwise). It is by applying this projection that one obtains a $\mathrm{II}_1$ algebra, as can be verified by computing the trace of the identity.

Elements of this algebra are generated by operators of the form
\be
\widehat{\sfa} = \Theta \, \e^{\ii p H} \, \sfa \, \e^{- \ii p H} \, \Theta \, ,
\ee
and bounded functions of $q$. Here $p$ is the variable conjugate to $q$. To uniformize with the notation of \cite{Chandrasekaran:2022cip}, we introduce the variable $x = -q$. If one conjugates the algebra by $\e^{- \ii p H}$, one gets an equivalent, perhaps simpler, description: the algebra $\cA$ is generated by $\sfa \in \cA_0$ and $H+x$, appropriately projected by $\Theta (q) = \Theta (-H -x)$.

While the algebra $\cA_0$ is a type $\rm{III}$ algebra, $\cA$ is a type $\rm{II}_1$ factor. It is also an ``hyperfinite'' algebra, meaning that it can be approximated by finite-dimensional matrix algebras. In the classification of von Neumann factors there exists only one hyperfinite type $\mathrm{II}_1$ factor up to isomorphisms. This factor is usually called $\cR$ in the literature and we will also adopt this notation. Note that for infinite dimensional algebras being isomorphic is a rather weak condition.

For a type $\rm{II}_1$ one can define a unique trace, up to normalization. This means that one can define density matrices and their von Neumann entropy. This entropy should be thought of as a renormalized entropy where an infinte constant, corresponding to the infinite entanglement of the vacuum state in quantum field theory, has been subtracted. The algebra $\cA$ has a state with maximum entropy, the so called tracial state, whose density matrix is the identity. The trace is defined as 
\be
\Tr \, \widehat{\sfa} = \braket{\Psi_{\mathrm{max}} \vert \widehat{\sfa} \vert \mathrm{\Psi}_{\mathrm{max}} }
\ee
where $\Psi_{\mathrm{max}} = \Psi_{\mathrm{dS}} \otimes \e^{-\beta_{\mathrm{dS}} q / 2} \sqrt{\beta_{\mathrm{dS}}}$. The state with maximum entropy is empty de Sitter space tensored with a state where the observer's energy has a thermal distribution. Indeed cyclicity of the trace follows from the thermal nature of de Sitter space: for 
\be
\Tr \, \widehat{\sfa} (t) \, \widehat{\sfb} = \Tr \, \widehat{\sfb} \, \widehat{\sfa} (t)
\ee
to hold, it has to be
\be
\braket{\sfb (0) \sfa (t)}_\beta = \braket{\sfa (t - \ii \beta) \sfb (0)}_\beta \, ,
\ee
which is the KMS condition. Here $\widehat{\sfa} (t) = \e^{\ii H t} \, \widehat{\sfa} \, \e^{- \ii H t}$ denote the usual time dependence of the operators.


 A particularly important class of states are the semiclassical states of the form $\widehat{\Phi} = \Phi \otimes f (x)$. Here $\Phi \in \cH_0$ and $f (x) \in L^2 (\real_+)$. For these states spacetime has a semiclassical character, where the observer can measure time with uncertainty in time smaller than $\beta_{\rm dS}$. To ensure this one can pick $f (x) = \sqrt{\varepsilon} \, g (\varepsilon x)$ with $\varepsilon << \beta_{\rm dS}$ and $g$ bounded, smooth and with support only for $x < 0$.
 
The density matrix associated with such a state is 
\be \label{rhoPhi}
\rho_{\widehat{\Phi}} = \frac{1}{\beta} \vert f \left( x + \frac{h_\Psi}{\beta} \right) \vert^2 \e^{- \beta x} \Delta_{\Phi \vert \Psi} + \cO (\varepsilon) \, .
\ee
Its entropy is the expectation value of the observable $S = - \log \rho_{\widehat{\Phi}}$
\begin{align} \label{logrho}
S (\rho_{\widehat{\Phi}}) & = - \braket{\widehat{\Phi} \vert \log \rho_{\widehat{\Phi}} \vert \widehat{\Phi}} \cr
& = - \braket{\Phi \vert h_{\Psi \vert \Phi} \vert \Phi} + \braket{\widehat{\Phi} \vert h_\Psi + \beta x \vert \widehat{\Phi}} + \int_{- \infty}^0 \dd x \vert f(x) \vert^2 \left( - \log \vert f(x) \vert^2 + \log \beta \right)
\end{align}
where $h_\Psi = \beta_{\mathrm{dS}} H$. This entropy is physically interpreted as the generalized entropy of the bifurcate horizon
\be
S_{\mathrm{gen} } = \frac{A}{4 G_N} + S_{\mathrm{out}} \, ,
\ee
where $A$ is the area of the horizon and $S_{\mathrm{out}}$ the entropy of the quantum fields outside the horizon. This entropy gives the entropy of the static patch in de Sitter. This entropy can also be interpreted as a relative entropy \cite{Longo:2022lod}.

\section{Fluctuation theorems}

In this section we will discuss fluctuation theorems. The strategies of the proofs are related to those of \cite{benoist} and \cite{strasberg}.  In \cite{benoist} the authors consider finite dimensional systems and then relate the two-time measurement process to the spectral measure of certain relative entropy operators in order to have a well defined thermodynamic limit. On the other hand \cite{strasberg} requires the system to be finite dimensional in order to use properties of pure states and the results only hold for particular choices of the weights of the final state. Our construction is different and uses the properties of type $\rm{II}_1$ factors.

\paragraph{Quantum nonequilibrium correlators.}

In ordinary quantum mechanics general aspects of nonequilibrium dynamics can often be captured by correlators of the form
\be
\Tr \left( \rho \, \Pi_\cO \, \Pi_{\cO'} (t) \right)
\ee
where $\Pi_\cO$ and $ \Pi_{\cO'} (t) $ are projections evaluated at different times and $\rho$ describes a state. Correlators of this form appear already in studying gravity in de Sitter space \cite{Susskind:2021omt}. A general discussion of correlators of this type together with several applications can be found for example in \cite{strasberg} and references therein.

In our case the analog of the above correlators are objects of the form
\be
\Tr \left( \rho_{\widehat{\Phi}_1}^\alpha \, \sigma_{\widehat{\Phi}_2}^{1-\alpha} \right) = \braket{\widehat{\Phi}_2 \vert \Delta^{\alpha}_{\rho_{\widehat{\Phi}_1} \vert \sigma_{\widehat{\Phi}_2}} \vert \widehat{\Phi}_2} 
\ee
where $\widehat{\Phi}_1$ and $\widehat{\Phi}_2$ are two semiclassical states and the density matrix $\rho_{\mathrm{max}}$ is used to define the trace. For future uses we have raised the two density operators to a certain power, but one could in general consider arbitrary functions of those operators.

In order to obtain concrete results we have to make some choices. To begin with we consider the case where one density matrix $\rho_{\widehat{\Phi}}$ is associated to a semiclassical state while the second corresponds to the same state evolved in time
\be
\tau \left( \rho_{\widehat{\Phi}} \right) = \e^{- \ii t H} \rho_{\widehat{\Phi}} \e^{\ii t H} \, .
\ee
In this case a natural observable is the entropy. Let $S = \int s \, \dd \sfe_s$ be the spectral resolution of the entropy observable $S = - \log \rho_{\widehat{\Phi}}$. Then
\begin{align}
\rho_{\widehat{\Phi}}^\alpha &= \int \e^{-\alpha s} \, \dd \sfe_s \, , \cr
\tau \left( \rho_{\widehat{\Phi}}^{1- \alpha} \right) &= \e^{- \ii t H} \rho_{\widehat{\Phi}} \int \e^{\alpha s} \,  \dd \sfe_s \, \e^{\ii t H} \, .
\end{align}
This implies that we can write the above correlators as
\be \label{Ren1}
\Tr  \left( \rho_{\widehat{\Phi}}^{\alpha} \ \tau ( \rho_{\widehat{\Phi}}^{1-\alpha}) \right)  =
\int \e^{- \alpha (s' - s)} \, \Tr \left(
\e^{-\ii t H} \rho_{\widehat{\Phi}} \, \dd \sfe_s \, \e^{\ii t H}  \dd \sfe_{s'} \,
\right) \ .
\ee

\paragraph{A remark on the spectral theorem.}  Above we have used the spectral theorem to find an explicit form of the correlator \eqref{Ren1}. However for the purpose of keeping this note short and to the point, it is useful to resort to a discrete description, even if approximate. This is mostly for notational convenience in order to simplify the mathematical jargon involved in the two-time measurement scheme in the following. However let us present a heuristic justification for this. Let $X$ be a generic observable. We decompose it as
 \be
 X = \int x  \, \dd \sfe_x = \int x \, \Pi (x) \dd x
 \ee
where we have interpreted the spectral measure as $\Pi (x) \dd x$, where $\Pi (x)$ is not a projection but a projection density, i.e. $\Pi (x) \Pi (x') = \delta (x-x') \Pi (x)$. Let us however assume that our observable is slowly varying and that our measure instrument has a finite accuracy $\Delta = x_{j+1} - x_j$. We can write
\be
X = \sum_j \int_{x_j}^{x_{j+1}} \, x \, \Pi (x) \dd x \sim  \sum_j \overline{x}_j \int_{x_j}^{x_{j+1}} \Pi(x) \dd x = \sum_j \overline{x}_j \Pi_j
\ee
where now $\Pi_j =\int_{x_j}^{x_{j+1}} \Pi(x) \dd x$ are standard projections $\Pi_i \Pi_j = \delta_{ij} \Pi_i$ and we have approximated the value of the observable $x$ with its average over the interval $\Delta$. In this case the spectral theorem has the same heuristic expression as  the spectral decomposition in finite dimensional quantum systems. In most of this note we will use this form of the spectral theorem, occasionally referring to the more correct expression.

Now we can summarize the above discussion about \eqref{Ren1} as
\begin{align} \label{relRenyi}
\Tr  \left( \rho_{\widehat{\Phi}}^{\alpha} \ \tau ( \rho_{\widehat{\Phi}}^{1-\alpha}) \right) 
&=
\sum_{s , s'} \e^{- \alpha (s' - s)} \, \Tr \left(
\e^{-\ii t H} \rho_{\widehat{\Phi}} \Pi_s \e^{\ii t H} \Pi_{s'}
\right)
\cr & = 
\sum_{s , s'} \e^{- \alpha (s' - s)} \,  
\braket{\Psi_{\mathrm{max}} 
\vert
\e^{-\ii t H} \rho_{\widehat{\Phi}} \Pi_s \e^{\ii t H} \Pi_{s'}
\vert
 \Psi_{\mathrm{max}}}
\, ,
\end{align}
where we have used the definition of the trace. The advantage of this expression will be apparent momentarily. It should be clear how to revert to the language of spectral measures; mathematically inclined readers are encouraged to do so.

\paragraph{Time reversal.} We will now assume that the particular states we are considering are invariant under time reversal. As discussed in \cite{Harlow:2018tng,Harlow:2023hjb,Susskind:2023rxm} in quantum gravity time reversal is a gauge symmetry. Furthermore in the presence of an observer it acts on the observer as well since she is entangled with the quantum fields. 

We assume the existence of an anti-linear involution $\vartheta$ which acts as $\vartheta \circ \tau^t = \tau^{-t} \circ \vartheta$. This involution acts on the operators via a unitary $U_{\vartheta}$ so that $\vartheta (\sfa) = U_{\vartheta} \sfa U_{\vartheta}^{-1}$. In particular for a time reversal invariant state $\omega$ we must have $\vartheta (\omega) = \omega$. A more detailed discussion of time invariance, close to our scope, is for example in the Appendix C of \cite{strasberg}.

%

Now in our case we assume that the Hamiltonian of the fields is time reversal invariant $\vartheta (H) = H$ and that the same holds for the semi-classical state $\rho_{\widehat{\Phi}}$. The assumption of time reversal invariance implies that
\be
\Tr \left(
 \rho_{\widehat{\Phi}}^{\alpha} \, \tau^t ( \rho_{\widehat{\Phi}}^{1 - \alpha} )
\right) 
= \Tr \, \vartheta \left(
 \rho_{\widehat{\Phi}}^{\alpha} \, \tau^t ( \rho_{\widehat{\Phi}}^{1 - \alpha} )
\right) 
=\Tr \left(
 \rho_{\widehat{\Phi}}^{\alpha} \, \tau^{-t} ( \rho_{\widehat{\Phi}}^{1 - \alpha} )
\right)  \, .
\ee
By using cyclicity of the trace we then find the identity
\be \label{TRidentity}
\Tr \left(
 \rho_{\widehat{\Phi}}^{\alpha} \, \tau ( \rho_{\widehat{\Phi}}^{1 - \alpha} )
\right) =\Tr \left(
 \rho_{\widehat{\Phi}}^{1-\alpha} \, \tau ( \rho_{\widehat{\Phi}}^{ \alpha} )
\right)
\ee
which we will use later on.

\paragraph{Measurements and Entropy Production}

Assume now that the system is in a semiclassical state $\rho_{\widehat{\Phi}}$ and that the observer is measuring the entropy observable $S = - \log \rho_{\widehat{\Phi}}$. Note that the maximum entropy state $\Psi_{\mathrm{max}}$ is not a semiclassical state in the sense of \cite{Chandrasekaran:2022cip}, and is therefore excluded from the following analysis. As before let $S = \sum_{s} s \, \Pi_s$ be the spectral decomposition of the entropy observable. 

Suppose the observer performs a measurement at time $t = 0$ and observes the eigenvalue $s$. After the measurement the state of the system is updated to
\be
\frac{\Pi_s \rho_{\widehat{\Phi}} \Pi_s}{\Tr \Pi_s \rho_{\widehat{\Phi}}} = \frac{\rho_{\widehat{\Phi}} \Pi_s}{\Tr \Pi_s \rho_{\widehat{\Phi}}} \, ,
\ee
since $\Pi_s$ is also in the spectral decomposition of $\rho_{\widehat{\Phi}}$. Here $p (s) = \Tr \, \Pi_s \rho_{\widehat{\Phi}}$ is the probability of observing $s$. This state evolves in time according to
\be
\tau^t \left( \frac{\rho_{\widehat{\Phi}} \Pi_s}{\Tr \Pi_s \rho_{\widehat{\Phi}}}
 \right) 
=
\e^{- \ii t H} \frac{\rho_{\widehat{\Phi}} \Pi_s}{\Tr \Pi_s \rho_{\widehat{\Phi}}}
\e^{\ii t H} \, .
\ee
The conditional probability of observing $s'$ at time $t > 0$ is then
\be
\Tr \left( \Pi_{s'} \ 
\e^{- \ii t H} \frac{\rho_{\widehat{\Phi}} \Pi_s}{\Tr \Pi_s \rho_{\widehat{\Phi}}}
\e^{\ii t H}
\right) \, .
\ee
Then the probability of observing both values is given by
\be
p (s' , s) = p (s' \vert s) p (s) = 
\Tr \left( \Pi_{s'} \ 
\e^{- \ii t H} \rho_{\widehat{\Phi}} \Pi_s \e^{\ii t H}
\right) \, .
\ee
We can ask what is therefore the probability of observing an average change of entropy $\overline{s} = \frac{s'-s}{t}$ in the time $t$
\begin{align}
\mathbf{P}_t (\overline{s}) 
& = 
 \sum_{s' , s } \delta \left( \left(s - s'\right) - t \overline{s} \right) 
\braket{\Psi_{\mathrm{max}} \vert
 \Pi_{s'} \ 
\e^{- \ii t H} \rho_{\widehat{\Phi}} \Pi_s
\vert \Psi_{\mathrm{max}}
}
\end{align}
where we have used the definition of the trace and the fact that $H \Psi_{\mathrm{max}} = 0$.

Note that it follows from \eqref{relRenyi} that
\be
\Tr \left(
 \rho_{\widehat{\Phi}}^{\alpha} \, \tau ( \rho_{\widehat{\Phi}}^{1 - \alpha} ) \right) = \sum_{\overline{s}} \mathbf{P}_t (\overline{s}) \e^{- t \alpha \overline{s}} \, .
\ee
Now under our assumptions of time-reversal invariance we can use the identity \eqref{TRidentity} to obtain
\be
\sum_{\overline{s}} \mathbf{P}_t (\overline{s}) \e^{- t \alpha \overline{s}} = \sum_{\overline{s}} \mathbf{P}_t (\overline{s}) \e^{-t (1 - \alpha) \overline{s}} \, ,
\ee
or equilvalently
\be
\sum_{\overline{s}} \left[ \mathbf{P}_t (\overline{s}) -  \mathbf{P}_t (-\overline{s}) \e^{t \overline{s}} \right] \e^{-t \alpha \overline{s}} = 0 \, .
\ee
Since this identity holds for arbitrary values of $\alpha$ and $t$ we conclude that
\be
\mathbf{P}_t (- \overline{s}) = \e^{- t \overline{s}} \mathbf{P}_t (\overline{s}) \, .
\ee
This is our first fluctuation theorem. Physically it implies that negative entropy fluctuations are exponentially suppressed respect to positive entropy fluctuations. Of course this is expected on physical grounds. Note however that no assumption of thermal equilibrium was made. This result is fully general and holds also outside of thermal equilibrium.


\paragraph{Remark.} The above equality is usually called the Jarzynski identity \cite{jarzynski}. In stochastic thermodynamics one writes
\be
\braket{\e^{-\overline{s}}} = \int \e^{-\overline{s}}  \mathbf{P}_t (\overline{s}) = \int \mathbf{P}_t (- \overline{s}) = 1 \, ,
\ee
since a probability distribution is normalized. If we write $\overline{s} = - \beta \Delta F + \beta W$, which expresses the entropy in terms of the free energy and work, this is precisely the standard form of the Jarzynski equality \cite{jarzynski}. We refer the reader to \cite{strasberg} and reference therein for a more complete discussion.

\paragraph{General Fluctuation Theorems.}


The general form of the fluctuation theorem comes by comparing a transition probability with the probability of the same process but time reversed, in a suitable sense. Consider a classical-quantum state $\widehat{\Phi}$ as before. Let now $Y$ be an arbitrary observable in $\cA$. Let $Y = \sum_y  y \, \Lambda_y$ be its spectral decomposition. The probability for the observer to measure the value $y_0$ at time $t=0$ and the value $y_\tau$ at time $t = \tau$ is
\be
P [y_\tau , y_0] = \Tr \left( \Lambda_{y_\tau} \e^{- \ii \tau H} \Lambda_{y_0} \, \rho_{\widehat{\Phi}} \, \Lambda_{y_0} \e^{\ii \tau H} \Lambda_{y_\tau} \right) \, .
\ee
We will consider now the ''time-reversed'' probability. By this we mean the following situation: we take as initial state the time evolved density  matrix $ \rho_{\widehat{\Phi}}^{tr} = \e^{- \ii \tau H}  \rho_{\widehat{\Phi}} \e^{\ii \tau H} $ and define the time-reversed evolution $\rho_{\widehat{\Phi}}^{tr}  (t) = \e^{\ii t H}  \rho_{\widehat{\Phi}} \e^{-\ii t H}$ so that $\rho_{\widehat{\Phi}}^{tr}  (\tau) = \rho_{\widehat{\Phi}}  (0)$. 

Reasoning as before, the corresponding two-times measurement statistics is given by the probability
\be
P^{tr} [y_0 , y_{\tau}] = \Tr \left( \Lambda_{y_0} \e^{ \ii \tau H} \Lambda_{y_\tau} \, \rho_{\widehat{\Phi}}^{\tr} \, \Lambda_{y_\tau} \e^{-\ii \tau H} \Lambda_{y_0} \right) \, .
\ee
Define now the quantity
\be
\Xi[y_\tau , y_0] = \log \frac{P [y_\tau , y_0]}{P^{tr} [y_0 , y_{\tau}] } = -  \Xi^{tr} [y_0 , y_\tau] \, .
\ee
By averaging over the probabilities and using that a probability distribution is normalized, we see immediately that
\be
\braket{\e^{-\Xi}} = \sum_{y_\tau , y_0} P [y_\tau , y_0] \e^{- \Xi [ y_\tau , y_0 ]} = 1
\ee
which, by using Jensen's inequality $\braket{\e^J} \ge e^{\braket{J}}$ implies $\braket{\Xi} \ge 0$.

If we define the probabilities
\begin{align}
p (\Xi) = \sum_{y_{\tau} , y_0} P [y_\tau , y_0] \, \delta \left( \Xi - \Xi[y_\tau , y_0]  \right) \, , \cr
p^{tr} (\Xi) = \sum_{y_{\tau} , y_0} P^{tr} [y_\tau , y_0] \, \delta \left( \Xi - \Xi^{tr} [y_0 , y_\tau]  \right) \, ,
\end{align}
we obtain the relation
\begin{align}
p (\Xi) &=  \sum_{y_{\tau} , y_0} P^{tr} [y_\tau , y_0]  \, \e^{\Xi [y_{\tau} , y_0 ]} \, \delta \left( \Xi - \Xi[y_\tau , y_0]  \right)  \cr
 & = \e^{\Xi} \sum_{y_{\tau} , y_0} P^{tr} [y_\tau , y_0]  \, \delta \left( \Xi  +  \Xi^{tr}[y_0 , y_\tau]  \right) \cr
 & = \e^{\Xi} p^{tr} (-\Xi) \, .
\end{align}

This expression is an \textit{abstract fluctuation theorem}. As it is stated it is purely formal. To find something more useful have to make some assumptions on the state $\rho_{\widehat{\Phi}}$ in order to express $\Xi$ in terms of physical quantitites. We assume we can write it as the following coarse-grained spectral decomposition 
\be
\rho_{\widehat{\Phi}} = \sum_y \frac{p_y}{d_y} \Lambda_y \, ,
\ee
where $p_y = \Tr \rho_{\widehat{\Phi}} \Lambda_y$ and $d_y$ is the Murray-von Neumann coupling constant, or dimension, of the projection, defined as $\Tr \Lambda_y = d_y$. Here $d_y \in [0,1]$ and can be interpreted as the dimension of the projection in a continuous sense, meaning that it measures how much of the identity the projection covers. We have introduced this number to ensure that $\Tr \rho_{\widehat{\Phi}} = 1$. We call this a coarse-grained spectral decomposition as the general projection may depend on other labels, which however here are summed over since they are not measured. This implies that $d_y$ can be thought of as the ``number of states'' for which the value $y$ is observed, even if this number is in general not an integer. Physically this is because the trace is appropriately renormalized to subtract an infinite constant. Note that these properties are characteristic of a type $\mathrm{II}_1$ algebra.

By using this assumption we find
\be
P [y_\tau , y_0] = \Tr \left( \Lambda_{y_\tau} \e^{- \ii \tau H} \Lambda_{y_0}  \sum_y \frac{p_y}{d_y} \Lambda_y \Lambda_{y_0} \e^{\ii \tau H} \Lambda_{y_\tau} \right) = \Tr \left( \Lambda_{y_\tau} \e^{- \ii \tau H} \Lambda_{y_0}  \e^{\ii \tau H}  \right)  \frac{p_{y_0}}{d_{y_0}} \, ,
\ee
where we have used the cyclicity of the trace and the properties of the projections. A similar computation holds for the time reversed process.
By comparing the probabilities of the forward and reversed process we find
\be
\log \frac{P [y_\tau , y_0]}{P^{tr} [y_0 , y_{\tau}] } = \log \frac{\Tr  \rho_{\widehat{\Phi}} \Lambda_{y_0}}{\Tr  \rho_{\widehat{\Phi}} \Lambda_{y_\tau}} \frac{d_{y_\tau}}{d_{y_0}} \, ,
\ee
where the universal terms $ \Tr \left( \Lambda_{y_\tau} \e^{- \ii \tau H} \Lambda_{y_0}  \e^{\ii \tau H}  \right) $ cancel out in the ratio due to the cyclicity of the trace. Note that $P [y_\tau , y_0]$ is essentially the probability of observing the transition from the state labelled by the value $y_0$ to the state labelled by $y_\tau$ in the time $\tau$. Then the above results is a quantum counterpart of the result in classical statistical mechanics that such a transition is proportional to the volume of phase space occupied by the initial coarse grained state. In this case what takes the place of the size of phase space is the projection acting on the Hilbert space.

Let us express more explicitly the probabilities $\Tr \rho_{\widehat{\Phi}} \Lambda$. To begin with we note that
\be
\Tr \rho_{\widehat{\Phi}} \Lambda \log \rho_{\widehat{\Phi}} \Lambda = \Tr \rho_{\widehat{\Phi}} \Lambda \log \rho_{\widehat{\Phi}}  
\ee
since $\Tr \rho_{\widehat{\Phi}} \Lambda \log \Lambda $ vanishes because a projection has eigenvalues zero or one. Then using \eqref{rhoPhi} and \eqref{logrho} we can write
\begin{align}
\Tr \rho_{\widehat{\Phi}} \Lambda \log \rho_{\widehat{\Phi}} \Lambda &= \Tr  \rho_{\widehat{\Phi}} \Lambda \left[ - h_{\Phi \vert \Psi} - \beta x + \log \vert f(x) \vert^2 - \log \beta \right]
\\
& = \Tr \rho_{\widehat{\Phi}} \Lambda h_{\Psi \vert \Phi} - \Tr \rho_{\widehat{\Phi}} \Lambda h_\Phi -\Tr \rho_{\widehat{\Phi}} \Lambda \left( \beta x + h_\Psi \right) + \Tr \rho_{\widehat{\Phi}} \Lambda \log \vert f(x) \vert^2 - \log \beta \Tr \rho_{\widehat{\Phi}} \Lambda \, , \nonumber
\end{align}
where we have used the identity $h_{\Phi \vert \Psi} = h_\Psi + h_\Phi - h_{\Psi \vert \Phi}$. Since $h_\Phi \ket{\Phi} = 0$ we arrive at
\be
\Tr \rho_{\widehat{\Phi}} \Lambda = \frac{1}{\log \beta} \left[ - S (\rho_{\widehat{\Phi}} \Lambda) + \Tr \rho_{\widehat{\Phi}} \Lambda h_{\Psi \vert \Phi}  -\Tr \rho_{\widehat{\Phi}} \Lambda \left( \beta x + h_\Psi \right) + \Tr \rho_{\widehat{\Phi}} \Lambda \log \vert f(x) \vert^2 \right] \, .
\ee

In this expression, if we neglect $\Lambda$, the terms on the right hand side are the von Neumann entropy, the relative entropy between the semiclassical state and the maximum entropy state, and the observer's energy. The last term combines with the left hand side to form the entropy of the fields exterior to the horizon as the horizon cut goes to future infinity. By reintroducing $\Lambda$ these terms maintain their significance, but the relevant state is now $\rho_{\widehat{\Phi}} \Lambda $, the semiclassical state where a particular value of the observable $Y$ was measured.

It is easy now to compute directly the ratio of the transition probabilities. We however follow another route, in order to find a more compact result. Let us introduce the normalized density matrix
\be
\rho_y = \frac{ \e^{-\mathscr{H}} \Lambda_y }{\Tr  \e^{-\mathscr{H}} \Lambda_y} \, ,
\ee
where we have defined $\mathscr{H} = - \log \rho_{\widehat{\Phi}}$. Then 
\be
S ( \frac{ \e^{-\mathscr{H}} \Lambda_y }{\Tr  \e^{-\mathscr{H}} \Lambda_y}) = \log \Tr \e^{-\mathscr{H}} \Lambda_y + \Tr \frac{ \e^{-\mathscr{H}} \Lambda_y }{\Tr  \e^{-\mathscr{H}} \Lambda_y} \mathscr{H} \, .
\ee
So we can write
\be
S (\rho_y ) = \log \Tr \rho_{\widehat{\Phi}} \Lambda_y + \Tr \rho_y \mathscr{H} \, .
\ee
Then we have
\begin{align}
\log \frac{P [y_\tau , y_0]}{P^{tr} [y_0 , y_{\tau}] } & = \log \frac{\Tr  \rho_{\widehat{\Phi}} \Lambda_{y_0}}{\Tr  \rho_{\widehat{\Phi}} \Lambda_{y_\tau}}  + \log \frac{d_{y_\tau}}{d_{y_0}} \cr
& = \left[ S (\rho_{y_0}) - S (\rho_{y_\tau}) \right] - \left(\Tr \rho_{y_0} \mathscr{H} - \Tr \rho_{y_\tau} \mathscr{H}  \right)  + \log \frac{d_{y_\tau}}{d_{y_0}} \, .
\end{align}
These terms have the following physical interpretations, which differ significantly from the standard quantum thermodynamics setup. The first term represents the difference in von Neumann entropy between the two states. Typically, this difference indicates which of the two processes is thermodynamically favoured. The second term represents the difference between the expectation values of the modular Hamiltonian associated with $\Phi$ in the two projected states. This is akin to the relation found in \cite{england}, which governs the energy balance in nonequilibrium self-replicating systems. The key difference here is that the modular Hamiltonian, rather than the ordinary Hamiltonian, is involved. 

As in standard quantum thermodynamics, equilibrium quantities—such as the modular Hamiltonian and entropy—determine the probability of a process, even when it is out of equilibrium. The physical interpretation, similar to that in \cite{england}, suggests that the second term can counterbalance the entropy change indicated by the first term. Consequently, a process that would otherwise be disfavored due to entropy considerations might occur with higher probability because of the shift in the modular Hamiltonian. 

The last term is unique to the structure of type $\mathrm{II}_1$ algebras: since the projection dimensions $d_{y}$ can be arbitrarily close to zero, this term can potentially dominate the first two. Note that this effect is a direct consequence of having incorporated the observer in order to gravitationally dress the physical observables and would not have been present otherwise.
Therefore we predict that in this setup there are processes that are entropically suppressed but can still be favoured due to this offset. Investigating the physical implications of this result would be very interesting.

\section{Quantum channels and subfactors}

In this Section, we describe the relationship between quantum channels and subfactors. A quantum channel is a map that transforms density matrices into density matrices and represents the most general set of manipulations that an observer can perform on a density matrix. 

In \cite{Longo:2017aet}, the relationship between quantum channels and von Neumann algebras was explored in detail. The author explained how a quantum channel is associated with a subfactor (or, more generally, a specific bimodule) and used modular theory to investigate certain thermodynamic properties. In particular, general results were established concerning the positivity of entropy, along with explicit expressions for the free energy. This construction is quite general and implies some of our results. 

However, our discussion in this Section takes a different approach from \cite{Longo:2017aet}. Instead of following their method, we will use the specific structure of type $\mathrm{II}_1$ factors as introduced in \cite{Jones}. Our aim is to take some steps towards providing a physical interpretation of the structures revealed in \cite{Jones} within the context of quantum channels. We will now construct a family of quantum channels using Jones' theory of subfactors \cite{Jones}.


In the Heisenberg picture, it is more natural to view a quantum channel not as a map between states but as a map between algebras of observables. The usual picture can be recovered by taking the dual and considering states defined on these algebras \cite{preskill}. The process of mapping one density matrix to another, such as through a noisy channel, can be modeled as a map from one algebra of observables to another. We interpret a subalgebra as analogous to a subsystem, meaning that only a subset of observables can be measured. For instance it can happen that observables that are distinguishable in the algebra $\cA$ may not remain distinguishable when restricted to a subalgebra.

We will not consider the most general quantum channel. However, starting with a von Neumann algebra, we can naturally construct quantum channels by examining subalgebras. Specifically, given a von Neumann algebra, there exists a canonical set of maps known as \textit{conditional expectations}, which are trace-preserving and completely positive, thereby corresponding physically to quantum channels. We have seen in \eqref{CEandK} how conditional expectations relate to Kraus operators and thus to quantum channels.

Consider a von Neumann factor $\cA$. We require our subalgebra to include the identity. Furthermore we want it to be a factor to enjoy the same causal properties of $\cA$. Therefore we require the map  $E : \cA \longrightarrow \cB$ to be a projection onto a subfactor $\mathcal{B} \subset \mathcal{A}$, that is
\be
E (\sfx) = \sfx \, , \qquad \forall \, \sfx \in \cB \, ,
\ee
which is also $\cB$-linear
\be
E \left( \sfx \, \sfa \, \sfy \right) = \sfx \, E (\sfa) \, \sfy \, , \qquad \forall \, \sfx, \sfy \in \cB \ \text{and} \ \forall \, \sfa \, \in \cA \, .
\ee
A $\cB$-linear projection which is also positive is called a conditional expectation. One can show that any conditional expectation is completely positive and is therefore a quantum channel (see for example Chapter 9 of \cite{Stratila}).

We will now specialize to the type $\mathrm{II}_1$ hyperfinite factor $\cR$ which governs the algebra of observables in de Sitter. Let $\cS \subset \cR$ a subfactor. Then there exists a unique conditional expectation $E$ compatible with the unique (up to rescaling) faithful normal trace. In other words specifying a conditional expectation on $\cR$ is equivalent to give a subfactor $\cS$; in this sense we can interpret every subfactor as a quantum channel.

Let us review a few basic facts about subfactors. If $\cS \subset \cR$ is a type $\mathrm{II}_1$ factor, the Jones index of $\cS$ in $\cR$ is 
\be
[\cR : \cS] = \dim_\cS L^2 (\cR) \, .
\ee
The index measures how much smaller is $\cS$ within $\cR$ and it is $\ge 1$ with equality iff $\cR$ and $\cS$ coincide. Since $\cR$ is hyperfinite, every subfactor with finite index is hyperfinite as well (and therefore isomorphic to $\cR$). In particular if $[\cR : \cS] < 4$, then $\cS$ is irreducible, that is $\cS' \cap \cR = \complex I$. Jones famously proved in \cite{Jones} that if $[\cR : \cS] < 4$, then
\be
[\cR : \cS] \in \left\{ 4 \cos^2 \left( \frac{\pi}{n+2} \right) \, : \, n = 1,2,3, \dots \right\} \, .
\ee
We refer the reader to \cite{Speicher} for detailed derivations of this and other claims that we will use in this note.

The conditional expectation $E \, : \, \cS \longrightarrow \cR$ is completely determined by the orthogonal projection $e_\cS \, : \, L^2 (\cR) \longrightarrow L^2 (\cS)$ in $\cB (L^2 (\cR))$, in the sense that for every $\sfx \in \cR$ we have
\be
E (\sfx) e_\cS = e_\cS \, \sfx \, e_\cS \, ,
\ee
which says that $E(\sfx)$ and  $e_\cS \, \sfx \, e_\cS$ agree on $L^2 (\cS)$. We can think of $L^2 (\cR)$ as the Hilbert space arising from the GNS construction, with a cyclic vector $\Omega$, so that $e_\cS$ projects onto the subspace $\cS \, \Omega$.

The subfactor $\cS$ is characterized by its basic construction. This works as follows. Define the algebra
\be
\cR_1 = \left\{ \cR \cup \left\{ e_\cS \right\} \right\}'' \subset \cB \left( L^2 (\cR) \right) \, .
\ee
The algebra $\cR_1$, usually denoted by $\braket{\cR , e_\cS}$ is the algebra generated by $\cR$ and by the projection $e_\cS$ in $\cB \left( L^2 (\cR) \right)$, and is called the basic construction for $\cS$. Since $\cS$ has finite index in $\cR$, then $\cR_1$ is a type $\mathrm{II}_1$ algebra which includes $\cR$ as a subfactor. Note however that it is not a ``new'' algebra, but a subalgebra of $\cB \left( L^2 (\cR) \right)$.

We can iterate this construction and define the Jones' tower of subfactors:
\be
\cS \subset \cR  \stackrel{e_{1}}{ \subset}\cR_1  \stackrel{e_{2}}{ \subset}\cR_2  \stackrel{e_{3}}{ \subset}\cR_3 \cdots 
\ee 
where each factor is defined inductively as $\cR_{i+1} = \braket{\cR_i , e_{i+1}}$, and $e_{i+1} \equiv e_{\cR_{i+1}} \, :  L^2 \left( \cR_{i+1} \right) \longrightarrow  L^2 \left(\cR_i \right)$ is the projection. The purpose of this construction is to fully characterize the subfactor $\cS$. While the first projection $e_\cS$ contains partial information, the whole tower is necessary to construct invariant objects that completely specify the subfactor. The sequence of larger algebras and projections refines the description of how $\cS$ sits inside $\cR$. Several sophisticated methods (such as the standard invariant and planar algebras) can be used to classify subfactors (see, for example, \cite{Speicher}). For our purposes, we only need to know that a subfactor, representing a quantum channel, is characterized by a series of maps. What is the physical meaning of these maps?

We propose the following interpretation: the tower of algebras $\cR_i$ and projections $e_i$ are needed to fully specify the subfactor $\cS$, i.e., the quantum channel type. A quantum channel can be seen as a generalized measurement where the outcome is not recorded \cite{preskill}. Thus, the objects needed to specify a channel are similar to those needed for a generalized measurement. In the finite-dimensional setting, an observer needs an auxiliary Hilbert space (like the ancilla discussed in Section \ref{QMT}) and extra operators (like the Kraus operators) to perform a generalized measurement. In other words, the observer requires additional structures beyond their Hilbert space. Similarly, new Hilbert spaces $L^2 (\cR_i)$ and projections $e_i$ between them are needed to fully specify the channel. Unlike the finite-dimensional case, where a single extra Hilbert space may suffice, the entire Jones tower is necessary here.

The overall picture is appealing: as in the finite-dimensional case without gravity, the observer requires additional structures to perform measurements or specify the quantum channel. The new operators correspond to more sophisticated models of the observer, equipped with increasingly advanced instruments. Each new algebra models the algebra of quantum fields along with the observer and an increasingly sophisticated experimental apparatus. Note that the algebra $\cR$ already represents the full algebra of observables in the static patch. In order to construct more refined quantum channels, the observer needs access to additional structures. This does not alter the physical picture: each new algebra can be found within $\cB (L^2 (\cR))$, and each new Hilbert space arises from a GNS construction. This means the observer does not need new fields or particles of an unknown type; the measurement instrument is simply an auxiliary sector constructed from already accessible physical objects.

We defer the problem of constructing a more precise model of an observer, which could reproduce the full Jones tower. Possibly the construction outlined in \cite{AliAhmad:2024wja,DeVuyst:2024pop,fewster,Hoehn:2023ehz} is the correct formalism to use. It would also be interesting to explore the general properties of quantum channels for gravitational systems using the formalism discussed above. We hope to report soon on these issues.
 
\section*{Acknowledgements} I am supported by GAST, and I am a member of INDAM and IGAP. Some of these results were presented at the International Congress of Mathematical Physics 2024, Strasbourg and at the Qubits and Spacetime Unit seminar at the Okinawa Institute of Science and Technology; I am thankful to the attendants for the useful feedback.

\end{document}